\documentclass[letter]{aa}
\usepackage{graphicx}
\usepackage{natbib}
\usepackage{caption}
\usepackage{subcaption}
\usepackage{multirow}
\usepackage{lineno}
\usepackage{txfonts}
\usepackage{amsmath}
\usepackage{hyperref}
\begin{document}


   \title{Helical flows along coronal loops following the launch of a coronal mass ejection}

   \titlerunning{Solar rollercoaster on 2024 May 14}
   \authorrunning{Nedal et al.}
   
   \author{Mohamed Nedal\inst{1}\fnmsep\thanks{Corresponding author: \email{mohamed.nedal@dias.ie}},
          David M.~Long\inst{2,1},
          Catherine Cuddy\inst{1,2},
          Lidia Van Driel-Gesztelyi\inst{3,4},
          \and
          Peter T.~Gallagher\inst{1}
          }

   \institute{Astronomy \& Astrophysics Section, School of Cosmic Physics, Dublin Institute for Advanced Studies, DIAS Dunsink Observatory, Dublin, D15 XR2R, Ireland.
        \and
             Centre for Astrophysics and Relativity, School of Physical Sciences, Dublin City University, Glasnevin Campus, Dublin, D09 V209, Ireland.
        \and
             Mullard Space Science Laboratory, University College London, Holmbury St Mary, Dorking, Surrey, RH5 6NT, UK.
        \and
            Konkoly Observatory, Research Centre for Astronomy and Earth Sciences, Hungarian Academy of Sciences, Konkoly Thege \'ut 15-17., H-1121, Budapest, Hungary.
             }



  \abstract
   {Solar flares and coronal mass ejections (CMEs) are manifestations of energy release in the solar atmosphere, which can be accompanied by dynamic mass motions and waves in the surrounding atmosphere.}
   {Here, we present observations of plasma moving in a helical trajectory along a set of coronal loops formed following the eruption of a  CME on 2024 May 14. This helical motion was observed in extreme ultraviolet (EUV) images from the Solar Dynamic Observatory (SDO), which provides new insights into plasma properties in a set of post-eruption coronal loops.}
   {We utilize images from the SDO Atmospheric Imaging Assembly (AIA) instrument to track the helical motion of plasma and to characterize its speed, acceleration, and physical properties. Additionally, we explore the evolution of the plasma density and temperature along the helical structure using the differential emission measure technique.}
   {The helical structure was visible in AIA for approximately 22 minutes, having a diameter of $\sim$22~Mm, and a total trajectory of nearly 184~Mm. Analysis of the AIA observations reveals that the plasma flow along this helical coronal loop exhibits speeds of 77--384~km s$^{-1}$ and temperatures ranging from 3.46 to 10.2~MK. Additionally, the densities were estimated to be between 4.3$\times$10$^6$ and 1.55$\times$10$^7$~cm$^{-3}$, with an estimated magnetic field strength of 0.05-–0.3~G.
}
   {Following the launch of a CME, we find clear evidence for impulsive heating and expansion of plasma that travels a helical trajectory along a set of post-eruption loops. These observations provide an insight into impulsive plasma flows along coronal loops and indeed the topology of coronal loops.}


   \maketitle
%

\section{Introduction}
    Coronal mass ejections (CMEs) are massive eruptions of plasma and magnetic fields from the Sun's corona that disturb the heliosphere and drive space weather phenomena at the Earth. Complex structures, such as loops and helical flux ropes, as well as phenomena resembling massive vortices or tornado-like formations, often accompany CME evolution, specifically in the solar corona \citep{Su_2013, Vourlidas_2014, Chen_2017, Devi_2021}. These whirling plasma formations, linked to twisted magnetic fields and magnetic reconnection, offer valuable insights into solar eruption mechanisms \citep{Huadong_2017, Xin_2017}.
    
    Solar tornadoes arise from swirling magnetic fields in the Sun’s atmosphere and are often linked to the barbs of solar prominences \citep{Wedemeyer_2013, Engvold_2015}. Up to 30 tornadoes may be active at any given time, particularly during solar maximum, serving as plasma sources or sinks for prominences \citep{Su_2012, Wedemeyer_2013}. Their predominantly vertical, helical magnetic fields indicate a dynamic relationship with prominences, as intermittent rotation may contribute to instability and eruptions \citep{Gonzalez_2016, Levens_2016}.
    
    Solar tornadoes are pivotal in supplying mass and twist to filaments, influencing their formation and eruptions \citep{Su_2012, Gunar_2023}. Magnetic twist, rotation, plasma-$\beta$, and viscosity significantly impact their dynamics, with magnetic twist dominant in coronal conditions and rotation more relevant in the photosphere \citep{Ghoraba_2018}.
    Despite the growing understanding of solar tornadoes, questions remain about their role in coronal heating and the detailed processes of magnetic reconnection \citep{Pontin_2012, Panesar_2013, Kuniyoshi_2024}. Comprehensive studies on solar tornado-like structures must consider their three-dimensional, non-uniform, and asymmetric evolution \citep{Su_2013, Schmieder_2017}.
    
    Here, we present a high-resolution observation of a helical mass motion formation in the solar corona, following a solar flare and CME eruption. To the best of our knowledge, this is the first time a helical flow has been clearly identifiable during the eruption of a CME event. This may offer new insights and constraints on CME models and mass-energy transport during eruptions. In Section 2, we describe the observations and data analysis techniques employed in this study. In Section 3, we present the results and their interpretation. Finally, in Section 4, we summarize our findings.

\section{Observations and data analysis}
    The Solar Dynamics Observatory (SDO)'s Atmospheric Imaging Assembly (AIA) provides continuous high-resolution imaging of the Sun's atmosphere using multi-wavelength channels \citep{SDO_2011, AIA_2012}. The extreme ultraviolet (EUV) channels in various ionized iron states allow for the construction of temperature maps of the solar corona, ranging from below 1 MK to above 20~MK. The 304~$\AA$ channel, which captures emissions from ionized helium (He~II), is particularly important for studying prominences, filaments, and chromospheric dynamics.
    AIA captures images up to 0.5~$R_\odot$ above the solar limb with a spatial resolution of about 1.5~arcseconds and a 12-second cadence, enabling precise tracking of dynamic phenomena like plasma flows and tornadoes. Its multi-wavelength imaging allows for multi-thermal diagnostics across various heights in the solar atmosphere.
    
\begin{figure*}[ht!]
    \centering
    \includegraphics[width=0.9\linewidth]{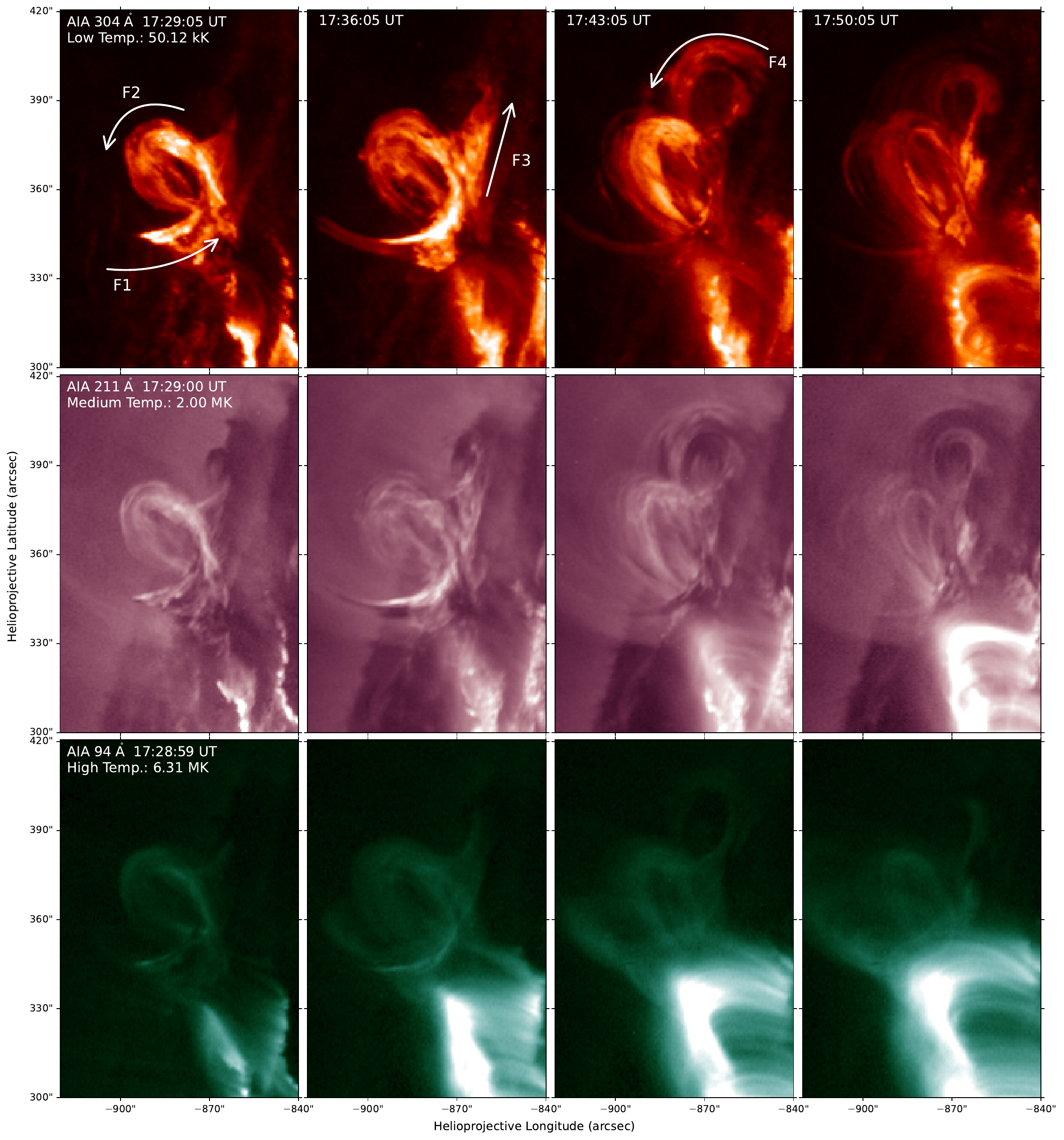}
    \caption{AIA images of the helical plasma in three channels that represent low (top row), medium (middle row), and high (bottom row) coronal temperatures. The white arrows in the top row show the direction of the plasma motion. Movies of the event are available in the online version.}
    \label{fig:aia_channels}
    \end{figure*}

    The plasma flow began forming at $\sim$17:10~UT during a CME event from the active region (AR)~13682 (E65N17). The active region had an area of 100~Mm$^2$, according to NOAA\footnote{National oceanic and atmospheric administration (NOAA)-Space weather prediction center (SWPC): \url{https://www.swpc.noaa.gov/products/solar-and-geophysical-activity-summary}}. The associated halo CME was detected in the solar and heliospheric observatory (SOHO)-Large angle and spectrometric coronagraph (LASCO)~C2 at 17:48:05~UT, with a linear speed of 1407~km~s$^{-1}$ and an acceleration of -40.3 m s$^{-2}$ as per the CDAW CME catalog\footnote{Coordinated data analysis workshops (CDAW): \url{https://cdaw.gsfc.nasa.gov/CME_list/}}. The CME was also associated with an M4.4-class flare (start: 17:25~UT, peak: 17:38~UT, end: 18:18~UT). Figure~\ref{fig:aia_channels} illustrates the helical evolution over four timesteps in three EUV channels, showing a twisted structure spanning $\sim$184 Mm.
    
\begin{figure*}[!htp]
    \centering
    \includegraphics[width=0.9\linewidth]{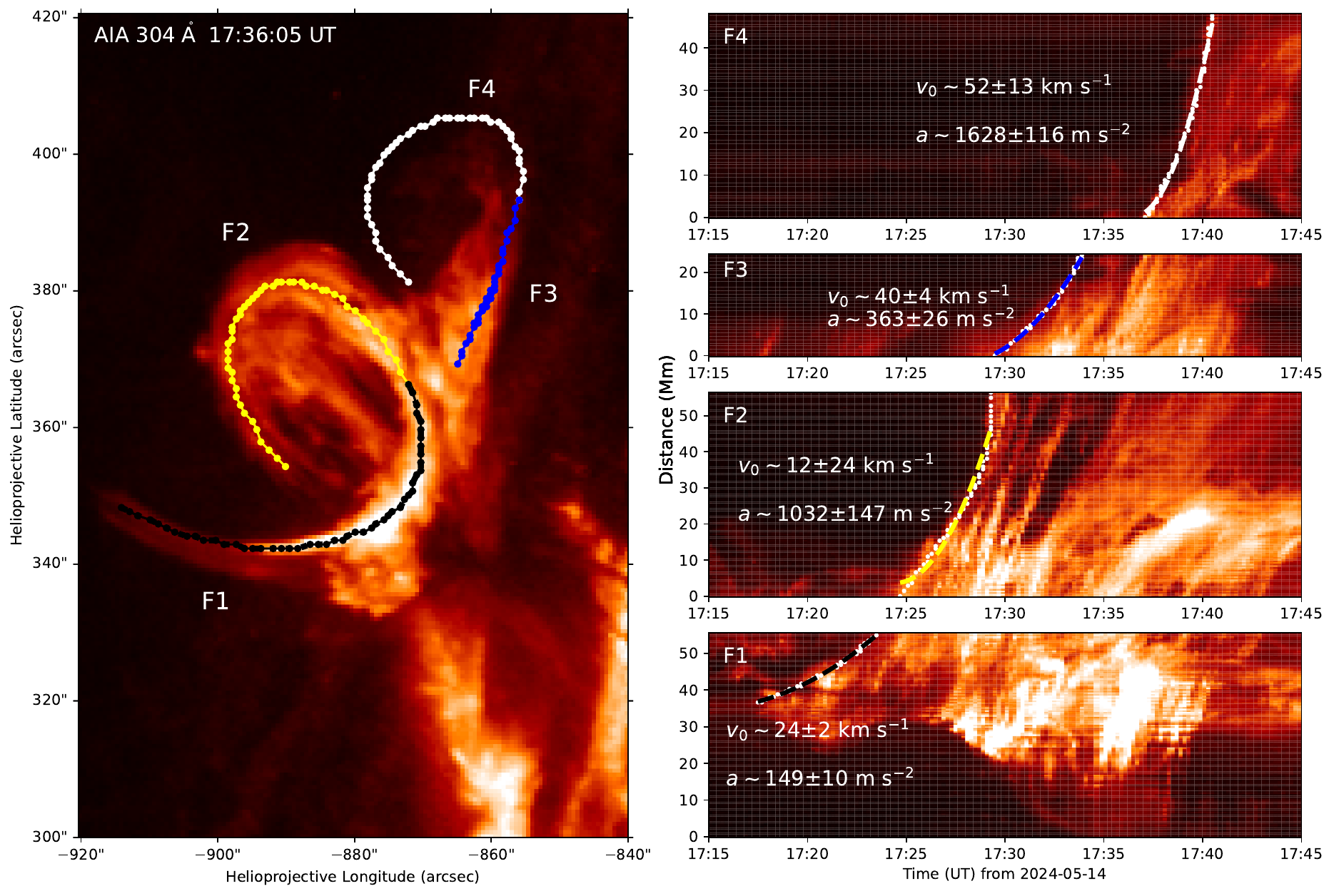}
    \caption{Time-dependent height analysis of four segments along the plasma helical structure. Left: snapshot image of the helical motion in four segments. Right: J-plots with the initial velocities (in km~s$^{-1}$) and acceleration (in m~s$^{-2}$) of the plasma in the four segments.}
    \label{fig:jmaps}
    \end{figure*}

    First, we obtained level-1 AIA data and used the aiapy Python package to upgrade the AIA data to level-1.5 and apply the PSF convolution. We tracked the evolution of the plasma flow along the structure in the 304~$\AA$ passband using B\'ezier curves \citep{mortenson_1999}, stacking the intensities with time to produce J-plots as shown in Figure~\ref{fig:jmaps}. By using a quadratic model to fit the bright structure in the J-plots, we estimated the projected speeds of the plasma motion in the four segments, starting from F1 up to F4, assuming constant acceleration. We calculated the initial position ($s_0$), initial velocity ($v_0$), acceleration ($a$), and instantaneous velocity ($v(t)$) from the curve fitting. We then computed the propagated uncertainties for minimum velocity ($v_{min}$), maximum velocity ($v_{max}$), and mean velocity ($v_{mean}$).
    The flow exhibited counterclockwise rotation at a projected velocity of $\sim$155$\pm$26~km~s$^{-1}$. These are projected speeds due to the flow’s position near the solar limb, which limits line-of-sight accuracy.
    
    Differential Emission Measure (DEM) analysis is a robust technique for studying coronal plasma temperature distributions. Various methods address its ill-posed inverse problem, including fast regularized inversion \citep{Plowman_2013}, enhanced algorithms with error estimates \citep{Hannah_2012}, and linear programming approaches \citep{Cheung_2015}. Recent advancements, such as a regularized maximum likelihood method \citep{Massa_2023}, enhance computational speed, noise robustness, and error estimation, enabling routine DEM map production for detailed studies of coronal thermal structures. Compared to earlier techniques \citep{Withbroe_1975, Sylwester_1980, Siarkowski_1983}, these methods provide more reliable insights into multi-thermal plasma environments.
    
    Despite its strengths, DEM analysis has limitations, including errors from noise and multi-thermality ambiguities that affect temperature resolution \citep{Guennou_2012}. Combining DEM analysis with solar rotational tomography \citep{Frazin_2005} has produced 3D coronal temperature maps. For instance, \citet{Sun_2014} studied an M7.7 flare, finding peak emission near the loop top at $\sim$16~MK. \citet{Levens_2015} applied DEM analysis to a solar tornado, revealing temperature-dependent velocity patterns and electron density distributions.
    
    Here, we used the regularized inversion of \citet{Hannah_2012} to investigate the temporal and spatial evolution of the plasma flow’s thermal structure. From the DEM results, we computed the average temperature ($T_{avg}$) and electron density ($n_e$) for each pixel using DEM-weighted averages. The line-of-sight depth, estimated from the observed plasma structure, was 2.24$\times10^{9}$~cm, based on the first plasma loop’s apparent diameter.
    
    Using $T_{avg}$ and $n_e$, we calculated the thermal pressure ($P_{th}$, in dyne~cm$^{-2}$) and sound speed ($C_s$, in km~s$^{-1}$). Assuming the plasma speed approximates the Alfv\'en speed ($v_A \approx v$), we inferred the magnetic field strength (in Gauss) in the flow’s four segments and estimated the magnetic pressure ($P_m$, in dyne~cm$^{-2}$). The plasma-$\beta$ parameter, representing the ratio of thermal to magnetic pressure, was calculated for the four segments of the helical flow. The minimum plasma-$\beta$ ranged from 0.89 to 18.19, while the maximum ranged from 1.87 to 119.13. The mean values for the segments were as follows: F1 (55.37), F2 (2.79), F3 (13.89), and F4 (1.33).
    The plasma flow’s four segments exhibited mean temperatures ranging from 2.46$\times10^6$~K to 4.74$\times10^6$~K and mean electron densities from 2.91$\times10^6$~cm$^{-3}$ to 7.93$\times10^6$~cm$^{-3}$.
    
    \begin{figure*}[ht!]
    \centering
    \includegraphics[width=0.9\linewidth]{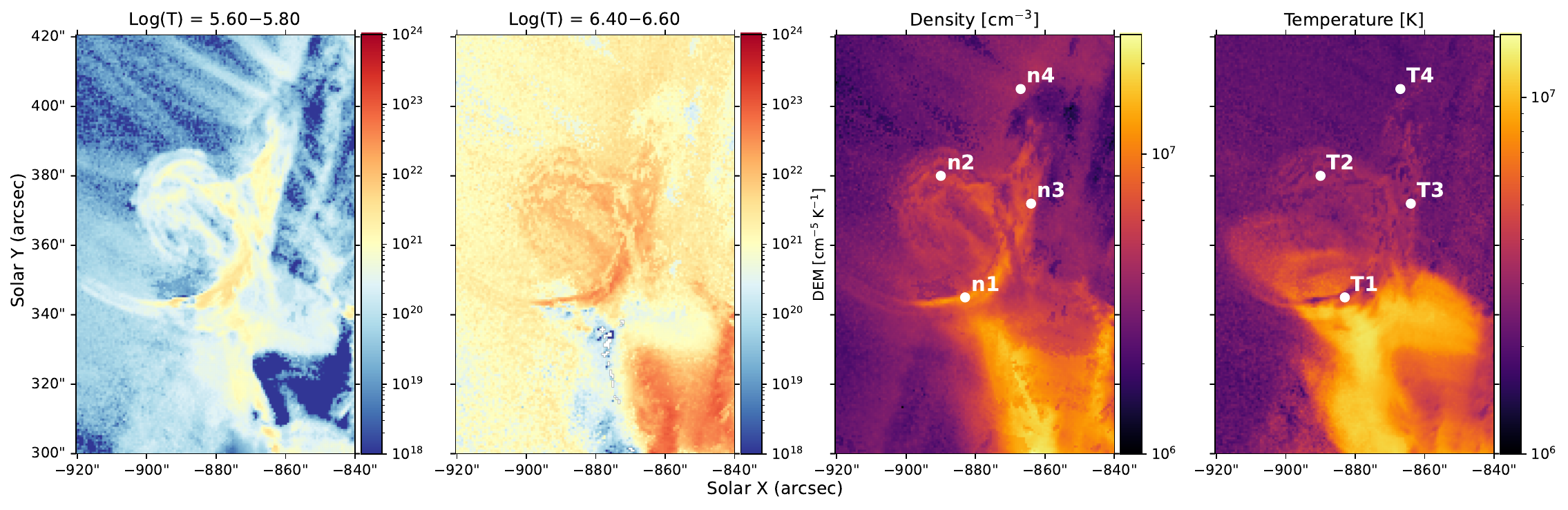}
    \caption{DEM analysis for the helical flow at a single time frame at 17:36~UT. Left panels: DEM output in two temperature ranges. Right panels: density and mean temperature deduced from the DEM analysis. Movies are available in the online version.}
    \label{fig:dem}
    \end{figure*}
    
    \begin{figure}[!htp]
        \centerline{
          \includegraphics[width=0.9\columnwidth]{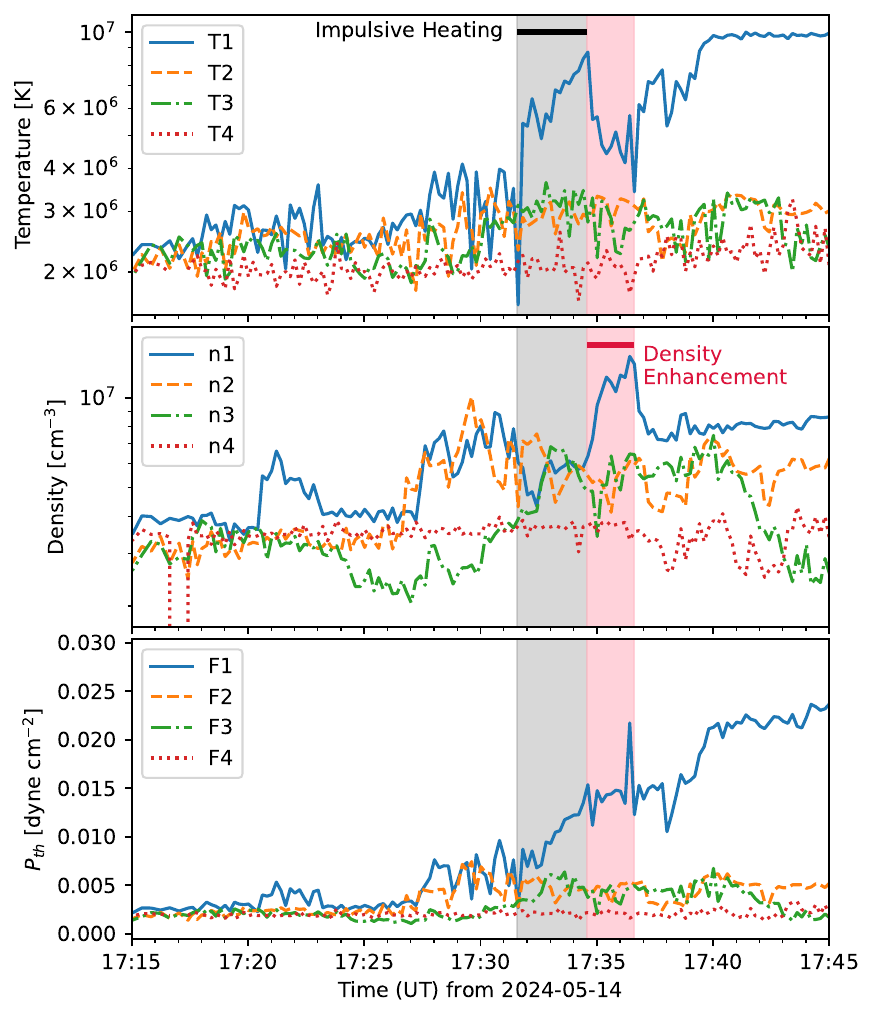}
          }
    \caption{Temporal evolution of key plasma parameters along the four segments of the helical structure. From top to bottom, we have the mean temperature, density, and thermal pressure at the four points (n1-n4 and T1-T4) along the plasma flow path in figure~\ref{fig:dem}.}
    \label{fig:temporal_evolution}
    \end{figure}

\section{Results and discussion}
The helical flow observed during the early stages of the CME eruption exhibited notable stability in size while undergoing significant rotational motion. Multi-wavelength observations revealed that the flow was closely tied to the underlying magnetic field structure, likely a segment of a flux rope. The rotational motion, evident in the AIA 304~$\AA$ observations, spanned between 17:17--17:41~UT, progressing through four distinct segments of the helical flow structure. J-maps constructed for these segments revealed spiraling plasma motions, with speeds varying significantly across the structure. The loops corresponding to segments F2 and F4 exhibited the highest speeds, ranging from 297 to 384~km~s$^{-1}$, while segments F1 and F3 displayed steadier, lower velocities, averaging 50–88~km~s$^{-1}$, indicative of more stable magnetic configurations. These variations likely reflect differences in local magnetic field configurations and reconnection dynamics.

The plasma dynamics and velocity distribution of the helical flow suggest a complex interaction between the local magnetic topology and energy release processes. DEM analysis showed that the helical plasma flow had a stratified thermal structure across temperature bins (Fig.~\ref{fig:dem}).
At lower temperatures ($log(T)=5.6-6.2$), the flow exhibited intricate structures with moderate DEM values, indicating condensed, cooler plasma. Intermediate temperatures ($log(T)=6.2-6.8$) showed enhanced DEM, particularly in the core of the helical flow, suggesting significant plasma heating. At higher temperatures ($log(T)=6.8-7.4$), the DEM became concentrated in the lower helical flow region, where flare ribbons and loops appeared, with values typical of flare-related heating above 10~MK. These trends align with scenarios of magnetic reconnection or shock heating during the event \citep{Longcope_2005, Aschwanden_2011, Schmelz_2013}.

The plasma density distribution further highlighted the helical plasma's stratified nature. Densities reached 10$^7$~cm$^{-3}$ in the flow's core, with steep gradients toward the periphery. The highest densities corresponded to localized areas of strong plasma emission, implying magnetic confinement or accumulation due to reconnection processes. The mean temperature map corroborated this, with the hottest regions overlapping the densest areas, consistent with localized energy deposition. These findings align with previous studies of solar eruptive structures, where plasma heating and energy release are concentrated in confined magnetic regions before dissipating \citep{Galeev_1981, Susino_2010, Reale_2014}.

The temporal evolution of thermal and magnetic pressures revealed distinct responses across the flow segments (Fig.~\ref{fig:temporal_evolution}). The thermal pressure dominated in F1 during impulsive heating, peaking around 17:34~UT, while the magnetic pressure played a stabilizing role in F2 and F4. Plasma-$\beta$ values further confirmed that thermal pressure dominated in F1 and F3 ($\beta$~$\gg$1), whereas F2 and F4 exhibited a balanced interplay between thermal and magnetic forces ($\beta$~$\sim$1).

Our results suggest that impulsive heating, triggered by the accompanying M-class flare, played a pivotal role in driving plasma evaporation and motion along pre-existing magnetic field structures, which is consistent with the findings of \citet{long_2023}. The rapid heating from $\sim$2~MK to $\sim$9~MK, followed by a density rise from $\sim$10$^6$~cm$^{-3}$ to $\sim$4~$\times$~10$^6$~cm$^{-3}$, highlights the role of flare energy release in plasma dynamics. This behavior parallels findings of explosive heating and chromospheric evaporation in low-density coronal plasma \citep{Bradshaw_2006}.

The observations suggest that the plasma predominantly filled pre-existing magnetic structures without significant evolution of the magnetic field itself. The flow's behavior differed from rapidly reforming flux ropes described by \citet{Miho_2014}. Instead, the observed plasma motion appears to result primarily from heating and overpressure effects induced by the eruption.

\section{Conclusions}
The helical mass motions observed following the CME eruption are an excellent demonstration of how complex formations interact with magnetic field structures and energy release processes. The flow's thermal stratification, density gradients, and velocity changes indicate how magnetic reconnection and impulsive heating shape its dynamics.

Key findings suggest that rapid plasma heating and evaporation, driven by the accompanying M-class flare, were integral to the helical flow's behavior. While the observed motion and plasma flows align with solar tornado characteristics, the lack of significant magnetic evolution and the plasma’s confinement to pre-existing structures point to a response dominated by CME-induced compression and overpressure effects rather than a fully developed tornado-like dynamic.

The plasma flow exhibited velocities ranging from 50 to 384~km~s$^{-1}$, with stratified thermal structure and densities peaking at 10$^7$~cm$^{-3}$. Thermal pressure dominated in some regions ($\beta \gg 1$), while others displayed a balance between thermal and magnetic forces ($\beta \sim 1$). The estimated magnetic field strength was significantly weaker than typical coronal loops, potentially due to projection effects, assumptions equating plasma velocity to the Alfv\'en speed, or DEM-based uncertainties in density estimation.
It is clear to us that the impulsive heating followed by density enhancement has yielded a pressure impulse in the flow, which is consistent with the work presented in the introduction.

These observations reinforce the role of complex flows as key elements in mass and energy transport during eruptive events. However, further studies, particularly those incorporating direct magnetic field measurements, are needed to fully unravel the interplay between plasma dynamics, magnetic reconnection, and CME evolution.

\begin{acknowledgements}
    Many thanks to Pascal Demoulin for the insightful discussions. Thank you to Shane Maloney, Paul Wright, Alasdair Wilson, and Laura Hayes for their helpful suggestions. Thank you to the referee for the constructive feedback. We acknowledge using the Python-curated version\footnote{\url{https://github.com/ianan/demreg}} of the regularized inversion of \citet{Hannah_2012} to calculate DEM from AIA observations. We acknowledge the use of data from the SDO/AIA instrument. This research used version 6.0.2\footnote{https://doi.org/10.5281/zenodo.13743565} of the SunPy open source software package \citep{sunpy_community2020}. This research used version 0.8.0\footnote{https://doi.org/10.5281/zenodo.5606094} of the aiapy open source software package \citep{Barnes2020}. This work also made use of Astropy:\footnote{http://www.astropy.org} a community-developed core Python package and an ecosystem of tools and resources for astronomy \citep{astropy:2013, astropy:2018, astropy:2022}. This work is supported by the project "The Origin and Evolution of Solar Energetic Particles”, funded by the European Office of Aerospace Research and Development under award No. FA8655-24-1-7392.
\end{acknowledgements}


\begin{thebibliography}{45}
	\expandafter\ifx\csname natexlab\endcsname\relax\def\natexlab#1{#1}\fi
	
	\bibitem[{Aschwanden \& Boerner(2011)}]{Aschwanden_2011}
	Aschwanden, M.~J. \& Boerner, P. 2011, \apj, 732, 81
	
	\bibitem[{{Astropy Collaboration} {et~al.}(2022){Astropy Collaboration}, {Price-Whelan}, {Lim}, {Earl}, {Starkman}, {Bradley}, {Shupe}, {Patil}, {Corrales}, {Brasseur}, {N{"o}the}, {Donath}, {Tollerud}, {Morris}, {Ginsburg}, {Vaher}, {Weaver}, {Tocknell}, {Jamieson}, {van Kerkwijk}, {Robitaille}, {Merry}, {Bachetti}, {G{"u}nther}, {Aldcroft}, {Alvarado-Montes}, {Archibald}, {B{'o}di}, {Bapat}, {Barentsen}, {Baz{'a}n}, {Biswas}, {Boquien}, {Burke}, {Cara}, {Cara}, {Conroy}, {Conseil}, {Craig}, {Cross}, {Cruz}, {D'Eugenio}, {Dencheva}, {Devillepoix}, {Dietrich}, {Eigenbrot}, {Erben}, {Ferreira}, {Foreman-Mackey}, {Fox}, {Freij}, {Garg}, {Geda}, {Glattly}, {Gondhalekar}, {Gordon}, {Grant}, {Greenfield}, {Groener}, {Guest}, {Gurovich}, {Handberg}, {Hart}, {Hatfield-Dodds}, {Homeier}, {Hosseinzadeh}, {Jenness}, {Jones}, {Joseph}, {Kalmbach}, {Karamehmetoglu}, {Ka{l}uszy{'n}ski}, {Kelley}, {Kern}, {Kerzendorf}, {Koch}, {Kulumani}, {Lee}, {Ly}, {Ma}, {MacBride}, {Maljaars}, {Muna}, {Murphy}, {Norman}, {O'Steen},
		{Oman}, {Pacifici}, {Pascual}, {Pascual-Granado}, {Patil}, {Perren}, {Pickering}, {Rastogi}, {Roulston}, {Ryan}, {Rykoff}, {Sabater}, {Sakurikar}, {Salgado}, {Sanghi}, {Saunders}, {Savchenko}, {Schwardt}, {Seifert-Eckert}, {Shih}, {Jain}, {Shukla}, {Sick}, {Simpson}, {Singanamalla}, {Singer}, {Singhal}, {Sinha}, {Sip{H{o}}cz}, {Spitler}, {Stansby}, {Streicher}, {{{S}}umak}, {Swinbank}, {Taranu}, {Tewary}, {Tremblay}, {Val-Borro}, {Van Kooten}, {Vasovi{'c}}, {Verma}, {de Miranda Cardoso}, {Williams}, {Wilson}, {Winkel}, {Wood-Vasey}, {Xue}, {Yoachim}, {Zhang}, {Zonca}, \& {Astropy Project Contributors}}]{astropy:2022}
	{Astropy Collaboration}, {Price-Whelan}, A.~M., {Lim}, P.~L., {et~al.} 2022, \apj, 935, 167
	
	\bibitem[{{Astropy Collaboration} {et~al.}(2018){Astropy Collaboration}, {Price-Whelan}, {Sip{\H{o}}cz}, {G{\"u}nther}, {Lim}, {Crawford}, {Conseil}, {Shupe}, {Craig}, {Dencheva}, {Ginsburg}, {Vand erPlas}, {Bradley}, {P{\'e}rez-Su{\'a}rez}, {de Val-Borro}, {Aldcroft}, {Cruz}, {Robitaille}, {Tollerud}, {Ardelean}, {Babej}, {Bach}, {Bachetti}, {Bakanov}, {Bamford}, {Barentsen}, {Barmby}, {Baumbach}, {Berry}, {Biscani}, {Boquien}, {Bostroem}, {Bouma}, {Brammer}, {Bray}, {Breytenbach}, {Buddelmeijer}, {Burke}, {Calderone}, {Cano Rodr{\'\i}guez}, {Cara}, {Cardoso}, {Cheedella}, {Copin}, {Corrales}, {Crichton}, {D'Avella}, {Deil}, {Depagne}, {Dietrich}, {Donath}, {Droettboom}, {Earl}, {Erben}, {Fabbro}, {Ferreira}, {Finethy}, {Fox}, {Garrison}, {Gibbons}, {Goldstein}, {Gommers}, {Greco}, {Greenfield}, {Groener}, {Grollier}, {Hagen}, {Hirst}, {Homeier}, {Horton}, {Hosseinzadeh}, {Hu}, {Hunkeler}, {Ivezi{\'c}}, {Jain}, {Jenness}, {Kanarek}, {Kendrew}, {Kern}, {Kerzendorf}, {Khvalko}, {King}, {Kirkby}, {Kulkarni},
		{Kumar}, {Lee}, {Lenz}, {Littlefair}, {Ma}, {Macleod}, {Mastropietro}, {McCully}, {Montagnac}, {Morris}, {Mueller}, {Mumford}, {Muna}, {Murphy}, {Nelson}, {Nguyen}, {Ninan}, {N{\"o}the}, {Ogaz}, {Oh}, {Parejko}, {Parley}, {Pascual}, {Patil}, {Patil}, {Plunkett}, {Prochaska}, {Rastogi}, {Reddy Janga}, {Sabater}, {Sakurikar}, {Seifert}, {Sherbert}, {Sherwood-Taylor}, {Shih}, {Sick}, {Silbiger}, {Singanamalla}, {Singer}, {Sladen}, {Sooley}, {Sornarajah}, {Streicher}, {Teuben}, {Thomas}, {Tremblay}, {Turner}, {Terr{\'o}n}, {van Kerkwijk}, {de la Vega}, {Watkins}, {Weaver}, {Whitmore}, {Woillez}, {Zabalza}, \& {Astropy Contributors}}]{astropy:2018}
	{Astropy Collaboration}, {Price-Whelan}, A.~M., {Sip{\H{o}}cz}, B.~M., {et~al.} 2018, \aj, 156, 123
	
	\bibitem[{{Astropy Collaboration} {et~al.}(2013){Astropy Collaboration}, {Robitaille}, {Tollerud}, {Greenfield}, {Droettboom}, {Bray}, {Aldcroft}, {Davis}, {Ginsburg}, {Price-Whelan}, {Kerzendorf}, {Conley}, {Crighton}, {Barbary}, {Muna}, {Ferguson}, {Grollier}, {Parikh}, {Nair}, {Unther}, {Deil}, {Woillez}, {Conseil}, {Kramer}, {Turner}, {Singer}, {Fox}, {Weaver}, {Zabalza}, {Edwards}, {Azalee Bostroem}, {Burke}, {Casey}, {Crawford}, {Dencheva}, {Ely}, {Jenness}, {Labrie}, {Lim}, {Pierfederici}, {Pontzen}, {Ptak}, {Refsdal}, {Servillat}, \& {Streicher}}]{astropy:2013}
	{Astropy Collaboration}, {Robitaille}, T.~P., {Tollerud}, E.~J., {et~al.} 2013, \aap, 558, A33
	
	\bibitem[{Barnes {et~al.}(2020)Barnes, Cheung, Bobra, Boerner, Chintzoglou, Leonard, Mumford, Padmanabhan, Shih, Shirman, Stansby, \& Wright}]{Barnes2020}
	Barnes, W.~T., Cheung, M. C.~M., Bobra, M.~G., {et~al.} 2020, Journal of Open Source Software, 5, 2801
	
	\bibitem[{{Bradshaw} \& {Cargill}(2006)}]{Bradshaw_2006}
	{Bradshaw}, S.~J. \& {Cargill}, P.~J. 2006, \aap, 458, 987
	
	\bibitem[{{Chen} {et~al.}(2017){Chen}, {Zhang}, {Ma}, {Yan}, \& {Xue}}]{Huadong_2017}
	{Chen}, H., {Zhang}, J., {Ma}, S., {Yan}, X., \& {Xue}, J. 2017, \apjl, 841, L13
	
	\bibitem[{{Chen}(2017)}]{Chen_2017}
	{Chen}, J. 2017, Physics of Plasmas, 24, 090501
	
	\bibitem[{{Cheng} {et~al.}(2017){Cheng}, {Guo}, \& {Ding}}]{Xin_2017}
	{Cheng}, X., {Guo}, Y., \& {Ding}, M. 2017, Science China Earth Sciences, 60, 1383
	
	\bibitem[{Cheung {et~al.}(2015)Cheung, Boerner, Schrijver, Testa, Chen, Peter, \& Malanushenko}]{Cheung_2015}
	Cheung, M. C.~M., Boerner, P., Schrijver, C.~J., {et~al.} 2015, \apj, 807, 143
	
	\bibitem[{{Devi, Pooja} {et~al.}(2021){Devi, Pooja}, {D\'emoulin, Pascal}, {Chandra, Ramesh}, {Joshi, Reetika}, {Schmieder, Brigitte}, \& {Joshi, Bhuwan}}]{Devi_2021}
	{Devi, Pooja}, {D\'emoulin, Pascal}, {Chandra, Ramesh}, {et~al.} 2021, A\&A, 647, A85
	
	\bibitem[{{Engvold}(2015)}]{Engvold_2015}
	{Engvold}, O. 2015, in Astrophysics and Space Science Library, Vol. 415, Solar Prominences, ed. J.-C. {Vial} \& O.~{Engvold}, 31
	
	\bibitem[{Frazin {et~al.}(2005)Frazin, Kamalabadi, \& Weber}]{Frazin_2005}
	Frazin, R.~A., Kamalabadi, F., \& Weber, M.~A. 2005, \apj, 628, 1070
	
	\bibitem[{{Galeev} {et~al.}(1981){Galeev}, {Rosner}, {Serio}, \& {Vaiana}}]{Galeev_1981}
	{Galeev}, A.~A., {Rosner}, R., {Serio}, S., \& {Vaiana}, G.~S. 1981, \apj, 243, 301
	
	\bibitem[{González {et~al.}(2016)González, Ramos, Arregui, Collados, Beck, \& de~la Cruz~Rodríguez}]{Gonzalez_2016}
	González, M. J.~M., Ramos, A.~A., Arregui, I., {et~al.} 2016, \apj, 825, 119
	
	\bibitem[{Guennou {et~al.}(2012)Guennou, Auch√®re, Soubri√©, Bocchialini, Parenti, \& Barbey}]{Guennou_2012}
	Guennou, C., Auch\'ere, F., Soubri\'e, E., {et~al.} 2012, The Astrophysical Journal Supplement Series, 203, 26
	
	\bibitem[{{Gun{\'a}r} {et~al.}(2023){Gun{\'a}r}, {Labrosse}, {Luna}, {Schmieder}, {Heinzel}, {Kucera}, {Levens}, {L{\'o}pez Ariste}, {Mackay}, \& {Zapi{\'o}r}}]{Gunar_2023}
	{Gun{\'a}r}, S., {Labrosse}, N., {Luna}, M., {et~al.} 2023, \ssr, 219, 33
	
	\bibitem[{{Hannah} \& {Kontar}(2013)}]{Hannah_2012}
	{Hannah}, I.~G. \& {Kontar}, E.~P. 2013, \aap, 553, A10
	
	\bibitem[{Janvier {et~al.}(2014)Janvier, Aulanier, Bommier, Schmieder, D√©moulin, \& Pariat}]{Miho_2014}
	Janvier, M., Aulanier, G., Bommier, V., {et~al.} 2014, \apj, 788, 60
	
	\bibitem[{Kuniyoshi {et~al.}(2024)Kuniyoshi, Bose, \& Yokoyama}]{Kuniyoshi_2024}
	Kuniyoshi, H., Bose, S., \& Yokoyama, T. 2024, \apjl, 969, L34
	
	\bibitem[{{Lemen} {et~al.}(2012){Lemen}, {Title}, {Akin}, {Boerner}, {Chou}, {Drake}, {Duncan}, {Edwards}, {Friedlaender}, {Heyman}, {Hurlburt}, {Katz}, {Kushner}, {Levay}, {Lindgren}, {Mathur}, {McFeaters}, {Mitchell}, {Rehse}, {Schrijver}, {Springer}, {Stern}, {Tarbell}, {Wuelser}, {Wolfson}, {Yanari}, {Bookbinder}, {Cheimets}, {Caldwell}, {Deluca}, {Gates}, {Golub}, {Park}, {Podgorski}, {Bush}, {Scherrer}, {Gummin}, {Smith}, {Auker}, {Jerram}, {Pool}, {Soufli}, {Windt}, {Beardsley}, {Clapp}, {Lang}, \& {Waltham}}]{AIA_2012}
	{Lemen}, J.~R., {Title}, A.~M., {Akin}, D.~J., {et~al.} 2012, \solphys, 275, 17
	
	\bibitem[{{Levens} {et~al.}(2015){Levens}, {Labrosse}, {Fletcher}, \& {Schmieder}}]{Levens_2015}
	{Levens}, P.~J., {Labrosse}, N., {Fletcher}, L., \& {Schmieder}, B. 2015, \aap, 582, A27
	
	\bibitem[{{Levens} {et~al.}(2016){Levens}, {Schmieder}, {Labrosse}, \& {L{\'o}pez Ariste}}]{Levens_2016}
	{Levens}, P.~J., {Schmieder}, B., {Labrosse}, N., \& {L{\'o}pez Ariste}, A. 2016, \apj, 818, 31
	
	\bibitem[{{Long} {et~al.}(2023){Long}, {Green}, {Pecora}, {Brooks}, {Strecker}, {Orozco-Su{\'a}rez}, {Hayes}, {Davies}, {Amerstorfer}, {Mierla}, {Lario}, {Berghmans}, {Zhukov}, \& {R{\"u}disser}}]{long_2023}
	{Long}, D.~M., {Green}, L.~M., {Pecora}, F., {et~al.} 2023, \apj, 955, 152
	
	\bibitem[{{Longcope}(2005)}]{Longcope_2005}
	{Longcope}, D.~W. 2005, Living Reviews in Solar Physics, 2, 7
	
	\bibitem[{{Massa} {et~al.}(2023){Massa}, {Emslie}, {Hannah}, \& {Kontar}}]{Massa_2023}
	{Massa}, P., {Emslie}, A.~G., {Hannah}, I.~G., \& {Kontar}, E.~P. 2023, \aap, 672, A120
	
	\bibitem[{Mortenson(1999)}]{mortenson_1999}
	Mortenson, M.~E. 1999, Mathematics for computer graphics applications (Industrial Press Inc.)
	
	\bibitem[{{Mozafari Ghoraba} {et~al.}(2018){Mozafari Ghoraba}, {Abedi}, {Vasheghani Farahani}, \& {Khorashadizadeh}}]{Ghoraba_2018}
	{Mozafari Ghoraba}, A., {Abedi}, A., {Vasheghani Farahani}, S., \& {Khorashadizadeh}, S.~M. 2018, \aap, 618, A82
	
	\bibitem[{{Panesar} {et~al.}(2013){Panesar}, {Innes}, {Tiwari}, \& {Low}}]{Panesar_2013}
	{Panesar}, N.~K., {Innes}, D.~E., {Tiwari}, S.~K., \& {Low}, B.~C. 2013, \aap, 549, A105
	
	\bibitem[{Pesnell {et~al.}(2012)Pesnell, Thompson, \& Chamberlin}]{SDO_2011}
	Pesnell, W.~D., Thompson, B.~J., \& Chamberlin, P. 2012, The solar dynamics observatory (SDO) (Springer)
	
	\bibitem[{Plowman {et~al.}(2013)Plowman, Kankelborg, \& Martens}]{Plowman_2013}
	Plowman, J., Kankelborg, C., \& Martens, P. 2013, \apj, 771, 2
	
	\bibitem[{{Pontin}(2012)}]{Pontin_2012}
	{Pontin}, D.~I. 2012, Philosophical Transactions of the Royal Society of London Series A, 370, 3169
	
	\bibitem[{{Reale}(2014)}]{Reale_2014}
	{Reale}, F. 2014, Living Reviews in Solar Physics, 11, 4
	
	\bibitem[{Schmelz {et~al.}(2013)Schmelz, Pathak, Jenkins, \& Worley}]{Schmelz_2013}
	Schmelz, J.~T., Pathak, S., Jenkins, B.~S., \& Worley, B.~T. 2013, \apj, 764, 53
	
	\bibitem[{{Schmieder} {et~al.}(2017){Schmieder}, {Zapi{\'o}r}, {L{\'o}pez Ariste}, {Levens}, {Labrosse}, \& {Gravet}}]{Schmieder_2017}
	{Schmieder}, B., {Zapi{\'o}r}, M., {L{\'o}pez Ariste}, A., {et~al.} 2017, \aap, 606, A30
	
	\bibitem[{{Siarkowski}(1983)}]{Siarkowski_1983}
	{Siarkowski}, M. 1983, \solphys, 84, 131
	
	\bibitem[{{Su} {et~al.}(2013){Su}, {Veronig}, {Holman}, {Dennis}, {Wang}, {Temmer}, \& {Gan}}]{Su_2013}
	{Su}, Y., {Veronig}, A.~M., {Holman}, G.~D., {et~al.} 2013, Nature Physics, 9, 489
	
	\bibitem[{Su {et~al.}(2012)Su, Wang, Veronig, Temmer, \& Gan}]{Su_2012}
	Su, Y., Wang, T., Veronig, A., Temmer, M., \& Gan, W. 2012, \apjl, 756, L41
	
	\bibitem[{Sun {et~al.}(2014)Sun, Cheng, \& Ding}]{Sun_2014}
	Sun, J.~Q., Cheng, X., \& Ding, M.~D. 2014, \apj, 786, 73
	
	\bibitem[{Susino {et~al.}(2010)Susino, Lanzafame, Lanza, \& Spadaro}]{Susino_2010}
	Susino, R., Lanzafame, A.~C., Lanza, A.~F., \& Spadaro, D. 2010, \apj, 709, 499
	
	\bibitem[{{Sylwester} {et~al.}(1980){Sylwester}, {Schrijver}, \& {Mewe}}]{Sylwester_1980}
	{Sylwester}, J., {Schrijver}, J., \& {Mewe}, R. 1980, \solphys, 67, 285
	
	\bibitem[{{The SunPy Community} {et~al.}(2020){The SunPy Community}, Barnes, Bobra, Christe, Freij, Hayes, Ireland, Mumford, Perez-Suarez, Ryan, Shih, Chanda, Glogowski, Hewett, Hughitt, Hill, Hiware, Inglis, Kirk, Konge, Mason, Maloney, Murray, Panda, Park, Pereira, Reardon, Savage, Sipőcz, Stansby, Jain, Taylor, Yadav, Rajul, \& Dang}]{sunpy_community2020}
	{The SunPy Community}, Barnes, W.~T., Bobra, M.~G., {et~al.} 2020, The Astrophysical Journal, 890, 68
	
	\bibitem[{{Vourlidas}(2014)}]{Vourlidas_2014}
	{Vourlidas}, A. 2014, Plasma Physics and Controlled Fusion, 56, 064001
	
	\bibitem[{{Wedemeyer} {et~al.}(2013){Wedemeyer}, {Scullion}, {Rouppe van der Voort}, {Bosnjak}, \& {Antolin}}]{Wedemeyer_2013}
	{Wedemeyer}, S., {Scullion}, E., {Rouppe van der Voort}, L., {Bosnjak}, A., \& {Antolin}, P. 2013, \apj, 774, 123
	
	\bibitem[{{Withbroe}(1975)}]{Withbroe_1975}
	{Withbroe}, G.~L. 1975, \solphys, 45, 301
	
\end{thebibliography}

\end{document}